\documentclass[a4paper,11pt]{article}
\usepackage[utf8]{inputenc}
\usepackage{color}
\usepackage{amsfonts}
\usepackage{amssymb}
\usepackage{graphics}
\usepackage{fancyhdr}
\usepackage{fancybox}
 \usepackage{float}
\usepackage{geometry}
\usepackage{minitoc}
 \usepackage{indentfirst}
 \usepackage{multicol}
 \usepackage[outerbars]{changebar}
 \usepackage{pifont,textcomp}
 \usepackage{amsfonts}
 \usepackage{amsmath}
 \usepackage{amscd}
 \usepackage{array}
\usepackage{multirow}
 \usepackage{mathrsfs}
 \usepackage{amssymb}
 \usepackage{amsthm}

\title{Position-dependent mass  in strong quantum gravitational  background fields. 
	   }
	   
\author{ {Lat\'evi Mohamed Lawson}
\space\\\\
 	${}^{1}\!$African Institute for Mathematical Sciences (AIMS) Ghana\\
 	Summerhill Estates, East Legon Hills, Santoe, Accra\\
 	P.O. Box LG DTD 20046, Legon, Accra, Ghana\\\\
 	${}^{2}\!$ Universit\'{e} de Lom\'{e}, Facult\'{e} des Sciences, D\'{e}partement de Physique,\\
 	Laboratoire de Physique des Mat\'{e}riaux et des Composants\\
 	à Semi-Conducteurs, 01 BP 1515 Lom\'{e}, Togo\\\\
 	latevi@aims.edu.gh   	
}

\begin{document}
\maketitle

\begin{abstract}
	
   	More recently, we have proposed a set of noncommutative space that describes the quantum gravity at the Planck scale [J. Phys. A: Math. Theor. 53, 115303 (2020)]. The interesting significant result we found is that, the generalized uncertainty principle induces a maximal measurable length of quantum gravity. This measurement revealed strong quantum gravitational effects at this scale and predicted a detection of gravity particles with low energies. In the present paper, to make  evidence  this prediction, we study in this space, the dynamics of  a particle with position-dependent mass (PDM) trapped in an infinite square well. We show that, by increasing the quantum gravitational effect, the PDM of the particle increases and induces deformations of the quantum energy levels. These deformations are more pronounced as one increases the quantum levels allowing, the particle to jump from one state to another with low energies and with high probability densities.

\end{abstract}

{\bf Keywords:}  Generalized Uncertainty Principle; Minimal and maximal lengths of quantum gravity; Position-dependent mass;  Quantum infinite  square well

\section{Introduction}
 Max  Planck's  length $l_p$ is conjectured to be the  smallest length of nature where  the  quantum gravity effects are sensible. It is about $l_p=10^{-35}m$ beyond   which  space-time distances cannot be resolved. 
  All quantum gravity theories, such as string theory \cite{2}  and   loop  quantum gravity \cite{3} particularly accounts for the existence  of a  minimal measurable length $\Delta x$ which is  proportional to the  Planck length $l_p$.  To theoretically realize this minimal length scale, a simple model  has been introduced,  the so-called  Generalized Uncertainty Principle (GUP) \cite{4,5,6,7,8,9,10,11,12,13,14,15,16,17,18} which
  is a gravitational correction to quantum mechanics. Moreover, to test this length in practice, many physicists believe that it  is not possible, because it needs extremely high energies for its  detection. For my point of  view as some other physicists \cite{19,20,21,22,23}, the measurement  of this length scale is somewhat difficult, but not impossible. To circumvent this problem of high energy requirement, we recently proposed in two  dimensions (2D) a set of noncommutative algebra \cite{1} that makes quantum gravitational  measurement stronger  at the Planck scale.  This measurement is manifested by the existence of maximal length scale
$l_{max}$ which is consistent with the one proposed by Perivolaropoulos \cite{23}.

 One of the biggest properties of gravity in general relativity (GR) is
 that, the gravitational field becomes stronger for heavy systems that curve
 the space, allowing the surronding light systems to fall down with low energies. As space-time is distorted in a gravitational field, Einstein has shown that relativistic effects such as time dilation and length contraction take effect. Such properties of gravity can also be manifested in quantum mechanics (QM). Thus, based on the latter  properties, one can postulate that, strong quantum gravitational fields can increase the mass of particles in a certain volume in such a way that, the variation of their masses can be measured. To theoretically realize this process, we consider a quantum system  whose mass changes with the  position. Such a system is well known  in the literature \cite{23,24,25,26,27,28,29} and is used to describe the dynamics of particles in semiconductor heterostructures materials \cite{30,31,32,33}. In the present situation,  we trap  this system in an infinite square well potential in which the quantum gravitational background fields are strong. Then, we study the spectrum of this system by solving analytically the  Schrödinger equation.   Firstly,  we observe that the states of the  system exhibit properties similar to Gaussian states of the  standard QM which are  consequences of quantum fluctuations  at this scale. Secondly,  
 we remark  that, the energy spectrum of this system  is  weakly proportional to the ordinary one of quantum mechanics free of the  gravitational  theory. This indicates that the effects of quantum gravity in this well increase the PDM which  consequently induces  deformations of energy levels so as to enable the particle to jump from one state into another with low energies and  with high  probability densities.

  This paper is organised  as follows: In the next  section, we  review in 1D, the outgoing of this GUP model with strong quantum gravity measurement and the corresponding  local translation generator. Section \ref{sec3} is devoted to the study of PDM in an infinite square-well potential in strong quantum gravitational  background fields. We give the  spectrum of this system by solving analytically the  Schrödinger equation. Finally, we conclude this work in section \ref{sec5}.

\section{  GUP with maximal length and its deformed translation symmetry } \label{sec3}
 In this section, we review the 1D GUP model with maximal length and {\color{red} its }
corresponding local translation generator. 
\subsection{ GUP with  maximal length }
 
 Let consider  a set of symmetric operators $\hat X, \hat Y, \hat P_x, \hat P_y$  defined on  the 2D Hilbert space   $\mathcal{H}_{\theta\tau}= \mathcal{L}^2(\mathbb{R}^2)$  which satisfy the following commutation relations and all possible permutations of the Jacobi identities \cite{1}
\begin{eqnarray} \label{alg1}
	[\hat X,\hat Y]&=&i\theta (\mathbb{I}-\tau\hat Y +\tau^2 \hat Y^2),\quad [\hat X,\hat P_x ]=i\hbar (\mathbb{I}-\tau \hat Y +\tau^2 \hat Y^2),\cr
	{[\hat Y,\hat P_y ]}&=&i\hbar (\mathbb{I}-\tau \hat Y +\tau^2 \hat Y^2),\quad\quad
	{ [\hat P_x,\hat P_y]}=0, \cr 
	{[\hat Y ,\hat P_x]}&=&0,\quad [\hat X,\hat P_y]=i\hbar\tau(2\tau \hat Y\hat X-\hat X)+
	i\theta\tau(2\tau \hat Y\hat P_y-\hat P_y),
\end{eqnarray}
where $\theta,\tau \in \mathbb{R}_+^*$  are both deformed parameters that describe  the Planck scale.
 The parameter $\tau$ is the GUP  deformed parameter \cite{6,7,8} related to  quantum gravitational effects   while 
the parameter $\theta$ is related to the    noncommutativity of the space  \cite{9,34} at this scale. In the framework of noncommutative classical or quantum mechanics, the parameter $\theta$  is proportional to the inverse of  constant magnetic   fields  $B$ (  $\theta\sim1/B$) \cite{35,36,37}. Since the algebra (\ref{alg1}) describes the space  at the  Planck scale, then such   magnefic  fields   are  necessarily  superstrong and  may play the  role of primordial magnetic fields. We  notice  that \cite{1} both   parameters are unified  as a minimal length scale $\Delta X_{min}=l_{min}=\tau\theta=\tau/B$. Futhermore,  this minimal length measure   $\Delta X_{min}$ is  coupled with the maximal length  $\Delta Y_{max}=l_{max}=\tau^ {-1}$ by the  inverse of strong magnetic fields $B$ as follows
\begin{eqnarray}
\Delta X\Delta Y=\frac{1}{B}= l_{min}l_{max}.
\end{eqnarray}
	At  the singulary point $l_{min}$ results of unification of magnetic  fields and quantum gravitational  fields. Since the Planck  length scale marks the frontier of our universe,  the origine of magnetic fields at this scale is   conjectured to   be  from  multiverse that  bounces  at this minimal position $l_{min}$.  As   predicted  in our previous  work \cite{1}, this scenario   breaks up the  big bang singularity. This  part of the paper is currently under investigation and  will be the object of further    exhaustive  study.   However, 
	simultaneous  measurements of operators $(\hat X,\hat P_x)$,  $(\hat X,\hat P_y)$, $(\hat P_x, \hat P_y)$ and  $(\hat Y,\hat P_x)$ are  spatial isotropy  since their measurements  do not present any minimal/maximal length or minimal momentum. Finally, for the case of interest in this paper i.e  a   simultaneous $\hat Y,\hat P_y$-measurement contains information on  how the properties of gravity  in QM  are similar to the ones of GR.

  The one dimensional set of the algebra (\ref{alg1}) becomes 
\begin{eqnarray}\label{alg2}
	{[\hat X,\hat P ]}=i\hbar (\mathbb{I}-\tau \hat X +\tau^2 \hat X^2),
\end{eqnarray}
where the operators $\hat X$ and $\hat P$ are defined on $  \mathcal{H}_{\tau}= \mathcal{L}^2(\mathbb{R})$. This  proposal (\ref{alg2})  follows the 
  recent approach of quantum gravity measurement introduced by Perivolaropoulos \cite{23}.   It has been shown that quantum gravity 
 naturally arises from cosmology due to the presence of particle horizons with the parameter $\alpha = \alpha_0 \frac{H_0^2}{c^2}$
  where $H_0$ is the Hubble constant, $c$ is the speed of light and $\alpha_0$ is a dimensionless parameter.
  
  From  the commutation relation  (\ref{alg2}),   an interesting feature  can be observed through the  following  uncertainty relation:
  \begin{eqnarray}\label{uncertitude}
  \Delta  X\Delta P\geq \frac{\hbar}{2}\left(1-\tau\langle \hat X\rangle+\tau^2\langle \hat X^2\rangle\right).\label{in2}
  \end{eqnarray}
  Using the relation 
  $ \langle \hat X^2\rangle=\Delta  X^2+\langle \hat X\rangle^2$,  equation (\ref{uncertitude}) can be rewritten as a second order equation for $\Delta X$. The solutions are 
  \begin{equation}
  \Delta X=\frac{\Delta P}{\hbar \tau^2}\pm \sqrt{\left(\frac{\Delta P}{\hbar \tau^2}\right)^2
  	-\frac{\langle \hat X\rangle}{\tau}\left(\tau\langle \hat X\rangle-1\right)
  	-\frac{1}{\tau^2}}.
  \end{equation}
  This equation leads to the absolute minimal uncertainty $\Delta P_{min}$ in $P$-direction  and the absolute maximal uncertainty  $\Delta X_{max}$ in $X$-direction  
  for $\langle  \hat X\rangle=0$
\begin{eqnarray}
 \quad \Delta X_{max}=l_{max}=\frac{1}{\tau}\quad  \mbox{and}\quad
 \Delta P_{min}=\hbar\tau.
\end{eqnarray}

It is well known  \cite{6} that, the   existence of minimal uncertainty   raised
the question of singularity of the space i.e the space is inevitably bounded by minimal quantity beyond which any further localization of particle is not possible. In the  presence situation, the minimal  momentum $\Delta P_{min}$   leads to the lost of representation in $P$-direction  whereas a maximal measurement $\Delta X_{max}$   is  the physical  space of  wavefunction representations i.e  all functions $\psi (X) \in  \mathcal{H}_{\tau}= \mathcal{L}^2(-\infty,+\infty)$   vanish  at the  boundary $ \psi(-\infty)=0=\psi (+\infty)$.

   From the uncertainty relation (\ref{uncertitude}),  it is interesting to observe that, the GUP is reduced into $ \Delta X \Delta P\sim \hbar $. As is well known from the Heisenberg principle, the latter relation can be casted into the form
 \begin{eqnarray}\label{E}
 	\Delta X\Delta E\sim \hbar c\implies \Delta E\sim \frac{\hbar c}{ 	\Delta X},
 \end{eqnarray}
 where  $\Delta E\sim \Delta P c$. Unlike the results obtained in \cite{4,11,12,13,14}, here  the required uncertainty energy  is  weak since the  dimension of  the length $	\Delta X$  is very large. This indicates that  a maximal  localization of quantum gravity induces  weak energies   for its measurement.  Let us now consider the equation $\Delta X= \Delta t c$. Inserting this equation in (\ref{E}), one obtains
 \begin{eqnarray}
 	\Delta t\sim \frac{\hbar }{\Delta E}.
 \end{eqnarray}
Since, the uncertainty energy is low in this space due to  the  strong quantum gravitational effects, then its time $\Delta t$ strongly increases i.e the time  runs more  slowly in this  space. In comparison with the minimal length theories, the  concept of a strong quantum  gravity  developed in this paper admits a close  analogy with the properties of gravity in GR  whereas the scenario of minimal length describes  an  exotic space i.e  a space  with low  quantum gravity  for which its properties  are out of the  common  sense.
 
 	One   recovers the algebra  (\ref{alg2}) by setting  the  operators  as follows 
	\begin{eqnarray}\label{reR}
	\hat X=\hat x,\quad 
	\hat P=(\mathbb{I}-\tau \hat x+\tau^2\hat x^2)\hat p,
	\end{eqnarray}
	where the symmetric operators $\hat x$ and $\hat p$  defined on  $  \mathcal{H}= \mathcal{L}^2(\mathbb{R})$ satisfy the Heisenberg algebra
	\begin{eqnarray}
	[\hat x,\hat p]=i\hbar.
	\end{eqnarray}
 From the representation (\ref{reR}), it follows  that  the position operator  $\hat X$ is symmetric while the momentum operator $\hat P$  is not. Thus, one observes
 \begin{eqnarray}
 \quad \hat X^\dag= \hat X, \quad  \hat P^\dag= \hat P+i\hbar\tau(\mathbb{I}-2\tau \hat X).
 \end{eqnarray} 
 In order to map this operator into symmetrical one, some synonymously  concepts  introduced in the literature can be used  such as: the $PT$-symmetry \cite{38,39}, the quasi-Hermiticity \cite{40,41}, the  pseudo-Hermiticity \cite{42,43,44}. In the presente case, to guarantee the  symmetry of the  operator $\hat P$, we propose  the following deformed completeness relation	
 
 \begin{eqnarray} \label{id} 
 \int_{-\infty}^{+\infty}\frac{dx}{1-\tau  x+\tau^2 x^2} |x\rangle\langle  x|=\mathbb{I}.
 \end{eqnarray}
 By performing a partial integration, one can easily show that \cite{1}
 \begin{eqnarray}
\langle \phi|\hat P\psi\rangle = \langle \hat P^\dagger \phi|\hat P\psi\rangle.
 \end{eqnarray}

Consequently, the scalar product between two states $|\psi\rangle$ and $|\phi\rangle$ and the orthogonality of
position eigenstate become
\begin{eqnarray}
		\langle \psi |\phi\rangle&=& \int_{-\infty}^{+\infty}\frac{dx}{1-\tau  x+\tau^2 x^2} \psi^*(x)\phi(x),\\
	\langle x |x'\rangle&=& (1-\tau  x+\tau^2 x^2)\delta(x-x').
\end{eqnarray}

 \subsection{ Deformed  translation  symmetry}
 Einstein’s theory of gravity is based on the symmetry principle of invariance under  general coordinate transformations, seen as local space-time transformations generated by the local translation generators. In this background, the gravitational force can be  modelled and described from a geometric point of view in terms of the curvature of space-time \cite{22}. Therefore, in this framework, considering  the  configuration space wave function defined by   $\psi(x):=\langle x |\psi\rangle$,  the moment operator $\hat P$  acts on $ \psi(x)$ as differential operator such as 
\begin{eqnarray}
\hat P\psi(x)&=&-i\hbar D_x\psi(x),\\
\frac{i}{\hbar}\hat P\psi(x)&=&  D_x\psi(x)\label{1}
\end{eqnarray}
where    $D_x= (1-\tau  x+\tau^2  x^2)\partial_x$ is  the deformed partial derivative  which introduces asymmetrical effects  in the system. For the operator $ \hat P$, one can associate through the exponential map  a $\tau$-translation operator $\hat {\mathcal{T}}_\tau$  that induces  an infinitesimal distance $\delta a$ in this direction
 \begin{eqnarray}
 	\hat {\mathcal{T}}_\tau (\delta a)=\exp\left(\frac{i}{\hbar}\delta a\hat P\right),\quad \mbox{with}\quad      \delta a\geq 0.
 \end{eqnarray}
 Since $\delta a$
 is infinitesimal, we expand the operator $ \hat {\mathcal{T}}_\tau$  to  first order of  $\delta a$ as follows
 \begin{eqnarray}
 	\hat {\mathcal{T}}_\tau (\delta a)=\mathbb{I}+\frac{i}{\hbar}\delta a\hat P\quad \mbox{and}\quad \lim_{\delta a\rightarrow 0} \hat {\mathcal{T}}_\tau (\delta a)= \mathbb{I}.
 \end{eqnarray}
 Since the operators  $\hat P$ is   symmetric,  one can readly show  that the $\tau$-translation operator $\hat {\mathcal{T}}_\tau$ forms a  unitary group  
 \begin{eqnarray}
 	\hat {\mathcal{T}}_\tau^\dag (\delta a) \hat {\mathcal{T}}_\tau (\delta a)=\mathbb{I}=\hat {\mathcal{T}}_\tau(\delta a) \hat {\mathcal{T}}_\tau^\dag (\delta a)\quad \mbox{and}\quad 
 	\hat {\mathcal{T}}_\tau (0)= \mathbb{I},
 \end{eqnarray}
 which implies
 \begin{eqnarray}
 	\hat {\mathcal{T}}_\tau^\dag (\delta a)=\hat {\mathcal{T}}_\tau^{-1} (\delta a)= \hat {\mathcal{T}}_\tau (-\delta a) =\mathbb{I}-\frac{i}{\hbar}\delta a\hat P.
 \end{eqnarray}
 Consequently,   for two arbitrary states $|\psi'\rangle$ and $|\phi'\rangle$ obtained from the action of $\hat {\mathcal{T}_\tau}$ on the respective states  $|\psi\rangle$ and $|\phi\rangle$, such that
 \begin{eqnarray}
 	|\psi'\rangle= \hat {\mathcal{T}}_\tau(\delta a)|\psi\rangle,\quad |\phi'\rangle= \hat {\mathcal{T}}_\tau(\delta a)|\phi\rangle,
 \end{eqnarray}
 one can easily check that the inner product of both states   is $\tau$-translational invariant
 \begin{eqnarray}
 	\langle \phi'|\psi'\rangle= \langle \phi|\psi\rangle.
 \end{eqnarray}
 The  action of  $\hat {\mathcal{T}}_\tau$ on the  state $|x\rangle$, 
 translates this state by the amount of $\delta a$  to the right
 \begin{eqnarray}
 	\hat {\mathcal{T}}_\tau(\delta a)|x\rangle= |x-\delta a( 1-\tau x+\tau^2 x^2)\rangle.
 \end{eqnarray}
 From this, it also follows that 
 \begin{eqnarray}
 	\langle x|\hat {\mathcal{T}}_\tau(\delta a)= \langle x+\delta a( 1-\tau x+\tau^2 x^2)|.
 \end{eqnarray}
 An alternative way of seeing this result without relying on position eigenstates representation is by considering
 the following action
 \begin{eqnarray}
 	\langle x|\hat {\mathcal{T}}_\tau(\delta a)|\psi\rangle&=&\psi(x)+\delta aD_x \psi(x),\cr
 	&=&\psi\left(x+\delta a( 1-\tau x+\tau^2 x^2)\right).
 \end{eqnarray}
 If the states of particles are moved by the amount $\delta a_1$ and $\delta a_2$, overall they have been moved  by the amount $\delta a_1+\delta a_2$
 \begin{eqnarray}
 	\hat {\mathcal{T}}_\tau(\delta a_1)	\hat {\mathcal{T}}_\tau(\delta a_2)\psi(x)&=&\psi(x+\delta a_1( 1-\tau x+\tau^2 x^2)+\delta a_2( 1-\tau x+\tau^2 x^2))\cr
 	&=& \hat {\mathcal{T}}_\tau(\delta a_1+\delta a_2)\psi(x).
 \end{eqnarray}

 Let us consider $\hat H$,    the operator Hamiltonian of a system defined within this 
 space  by 
 \begin{eqnarray}\label{ham}
 	\hat H (\hat P,\hat X)=\frac{\hat P^2}{2m_0} +V(\hat X),
 \end{eqnarray}
 where V is the potential energy of the system.  For periodical deformed potential energy  
 \begin{eqnarray}
 	{\hat {\mathcal{T}}_\tau}^\dag(\delta a)V(\hat x){\hat {\mathcal{T}}_\tau}(\delta a)=V(\hat x+\delta a( 1-\tau \hat x+\tau^2 \hat x^2))=V(\hat x),
 \end{eqnarray}
 the entire Hamiltonian satisfies
 \begin{eqnarray}
 	[\hat H,\hat {\mathcal{T}}_\tau ]=0.
 \end{eqnarray}

\begin{figure}[htbp]
 	\resizebox{1\textwidth}{!}{
 		\includegraphics{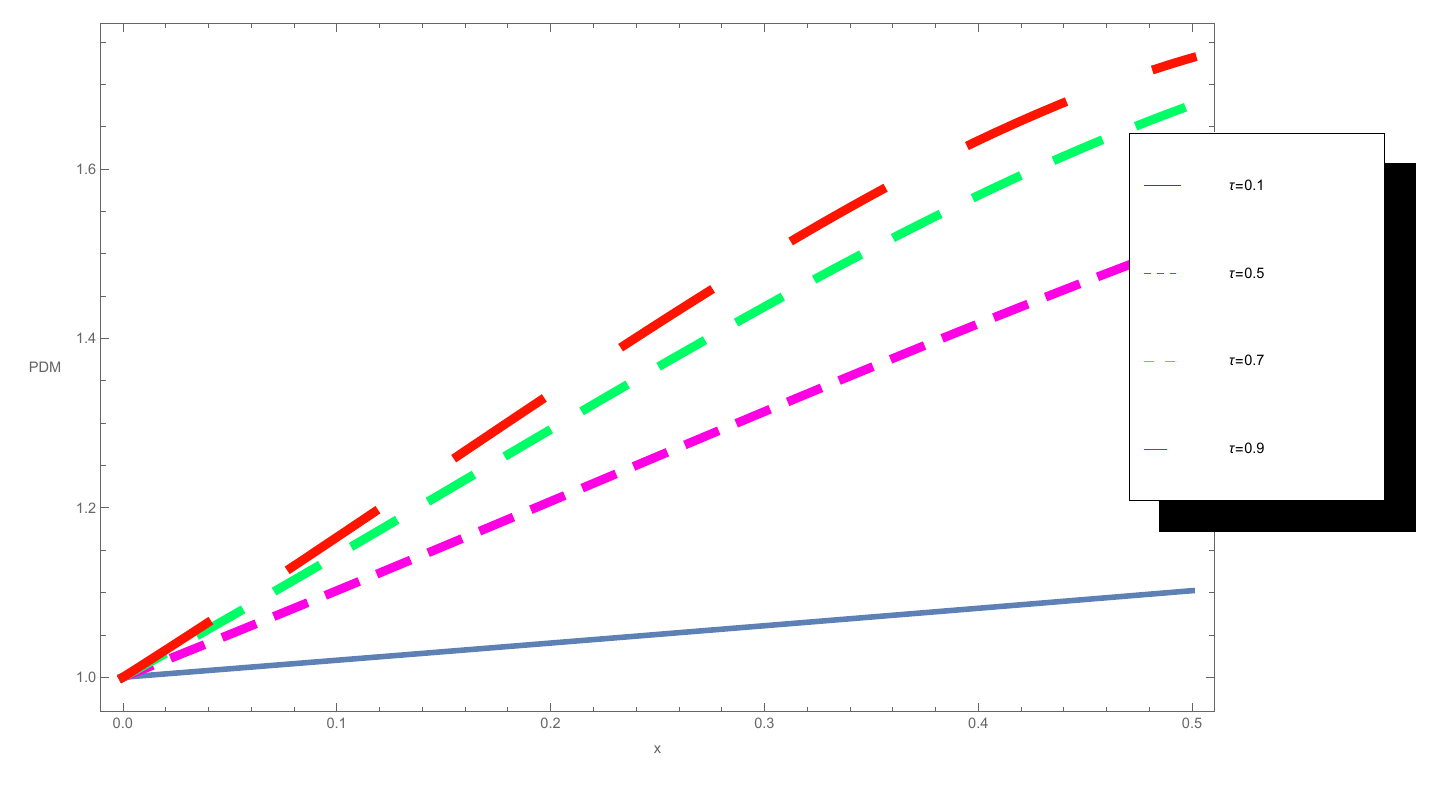}
 	}
 	\caption{\it \small PDM $m(x)$ versus the position x  for different values of $\tau$
 	}
 	\label{fig1}       
 \end{figure}

 The time-dependent deformed Schrödinger equation is 
 \begin{eqnarray}\label{sch}
 	-\frac{\hbar^2}{2m_0} D_x^2 \psi (x,t)+V(x,t)\psi (x,t)=i\hbar\partial_t\psi (  x,t).
 \end{eqnarray} 
 In analogy with problems involving PDM systems \cite{24,25,26,27,28,29,30}, this
 equation is rewritten as follows  
 \begin{eqnarray}
 i\hbar\partial_t\psi(x,t)=\left[-\frac{\hbar^2}{2m_0}\left(\sqrt{\frac{m_0}{m(x)}}\partial_x\right)^2 +V(x)\right]\psi(x,t)
 \end{eqnarray}
 where \begin{eqnarray}
 m(x)=\frac{m_0}{(1-\tau x+\tau^2 x^2)^2},
 \end{eqnarray}
  being the PDM's particle strongly  pertubated  by quantum gravity.  Figure \ref{fig1}  illustrates the PDM as a function of the  position $x\, \, (0<x<0.5)$. One  observes that   the effective mass $m(x)$ in this description increases with $\tau$. This indicates  that  quantum gravitational  fields increase with the effective $m(x)$ of the system. 
 In other words, by increasing  the mass of the  system or by just  taking   larger masses  and bringing them into  quantum  states, one can make the  quantum gravitational effects stronger for a measurement.  This   description  was  experimentally carried out  to measure  the
 gravitational field of significantly smaller source masses with remarkable  technical skills of realisation \cite{46}. Moreover, the  increase of $m(x)$ with the values of $\tau$ observed in this  system  is in  perfect analogy with the theory of  GR  where  massive objects    induce    strong gravitational  fields.  
  As   it will be shown in the  next section, the  increase of  the PDM of a particle in an infinite  square well potential  curves  the quantum levels, which  consequently implies  the decrease of  energy amplitudes and lead to  higher probability densities  of particles at this scale. 

 The probability density $\rho(x,t) =|\psi (x,t)|^2$ obeys the continuity equation
 \begin{eqnarray}
 	\frac{\partial \rho (x,t)}{\partial t}+  D_x J_{\tau}(  x,t)=0,
 \end{eqnarray}
 where the current density is given by  
 \begin{eqnarray} 	
 	J_{\tau}( x,t)=\frac{\hbar (1-\tau x+\tau^2 x^2)}{2im_0}\left( \psi^* \frac{d \psi}{dx}-\psi\frac{d\psi^*}{dx}\right).
 \end{eqnarray}

\section{PDM  particle in an infinite square well  }\label{sec3}

\subsection{ Eigensystems of  PDM  particle  }

Let us consider  the Hamiltonian of  the above  quantum system (\ref{ham}) confined in an infinite square-well potential, defined as

\begin{eqnarray}
V(x)=
\begin{cases}
0,\quad   0<x<L,  \\
\infty,\quad \mbox{  otherwise}.
\end{cases}
\end{eqnarray}
For standing waves in a null potential, the wave function $\psi(x)$
satisfying equation (\ref{sch}) obeys
\begin{eqnarray}\label{eq}
-\frac{\hbar^2}{2m_0} D_x^2 \psi (x)=E\psi (x), \quad \mbox{with} \quad E>0,
\end{eqnarray} 
	or 
	\begin{eqnarray}\label{eq1}
	-\frac{\hbar^2}{2m_0}\left[(1-\tau  x+\tau^2 x^2)^2\frac{d^2}{dx^2}- \tau(1-2\tau x)(1-\tau  x+\tau^2 x^2)\frac{d}{dx}\right] \psi (x)=E \psi (x).
	\end{eqnarray}
The   solution of the equation (\ref{eq}) is  given by
\begin{eqnarray}\label{wav}
\psi_k(x)= A\exp\left(i\frac{2k}{\tau\sqrt{3}}
\left[\arctan\left(\frac{2\tau x-1}{\sqrt{3}}\right)+\frac{\pi}{6}\right]\right),
\end{eqnarray}
where $k=\frac{\sqrt{2mE}}{\hbar} $. Then by normalization $\langle \psi_k|\psi_k\rangle=1$, we have 
\begin{eqnarray}
1&=&\int_{0}^{L}\frac{dx}{1-\tau x+\tau^2 x^2} \psi^* (x) \psi (x)\cr
&=& A^2 \int_{0}^{L}\frac{dx}{1-\tau  x+\tau^2 x^2}.
\end{eqnarray}
so, we find
\begin{eqnarray}\label{nzeta}
A=\sqrt{\frac{\tau\sqrt{3}}{2}} \left[\arctan\left(\frac{2\tau L-1}{\sqrt{3}}\right)+\frac{\pi}{6}\right]^{-1/2}.
\end{eqnarray}

\begin{figure}[htbp]
	\resizebox{0.8\textwidth}{!}{
		\includegraphics{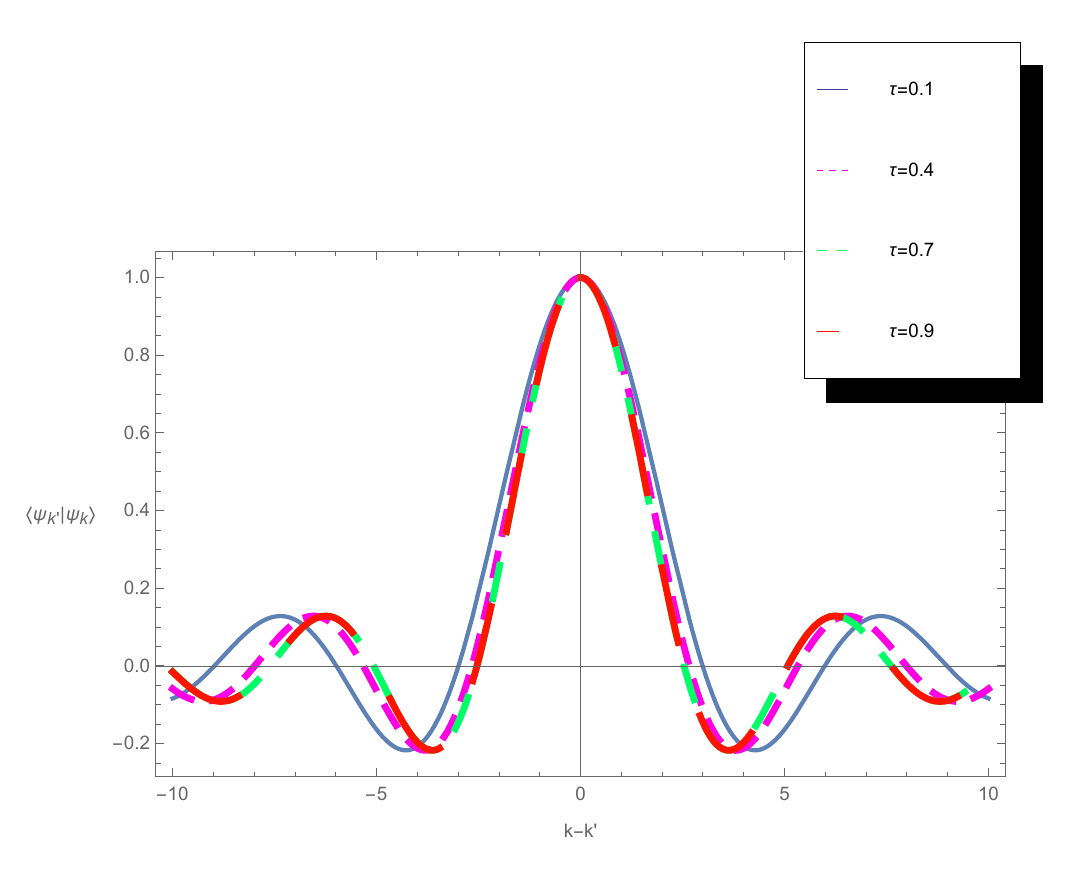}
	}
	\caption{\it \small Variation of $\langle \psi_{k'}|\psi_{k}\rangle $ versus $k-k'$ with $L=1$
	}
	\label{fig2}       
\end{figure}

 Based on the reference  \cite{7,46'}, the scalar product of the formal  eigenstates  is given by
\begin{eqnarray}
	\langle \psi_{k'}|\psi_k\rangle&=&\frac{\tau\sqrt{3}}{2(k-k')\left[\arctan\left(\frac{2\tau L-1}{\sqrt{3}}\right)\right]}\sin\left(\frac{2(k-k')\left[\arctan\left(\frac{2\tau L-1}{\sqrt{3}}\right)\right]}{\tau\sqrt{3}}\right).
\end{eqnarray}
This relation  shows that, the normalized  eigenstates (\ref{wav}) are  no longer
orthogonal. However,  if one tends $(k-k')\rightarrow \infty$, these states  become orthogonal 
\begin{eqnarray}
	\lim_{(k-k')\rightarrow \infty} \langle \psi_{k'}|\psi_{k}\rangle=0.
\end{eqnarray}
These properties show that, the states
$|\psi_k\rangle$ are essentially Gaussians centered at $(k-k')\rightarrow 0$ (see Figure \ref{fig2}). This observation indicates   quantum gravitational fluctuations at this scale. 

We suppose  that the wave function satisfies the  Dirichlet condition i.e it vanishes at the boundaries    $\psi(0)=0=\psi(L)$.
Thus, using especially the boundary condition $\psi(0)=0$,  the 
above wavefunctions (\ref{wav}) becomes
\begin{eqnarray}
\psi_k (x) =  A\sin\left(\frac{2k}{\tau \sqrt{3}}\left[\arctan\left(\frac{2\tau x-1}{\sqrt{3}}\right)
+\frac{\pi}{6}\right]\right).
\end{eqnarray}
The quantization follows from the boundary condition $\psi(L)=0$ and leads to the equation

\begin{eqnarray}
\frac{2k_n}{\tau \sqrt{3}}\left[\arctan\left(\frac{2\tau L-1}{\sqrt{3}}\right)
+\frac{\pi}{6}\right]&=&n\pi \quad \,\,\mbox{with}\quad n\in \mathbb{N}^*,\\
k_n &=&\frac{\pi\tau\sqrt{3}n}{2\left[\arctan\left(\frac{2\tau L-1}{\sqrt{3}}\right)
	+\frac{\pi}{6}\right]}.
\end{eqnarray}
Then, the  energy spectrum of the particle  is written as
\begin{eqnarray}\label{enrg1}
E_n &=& \frac{3\pi^2 \tau^2\hbar^2n^2}{8m_0 \left[\arctan\left(\frac{2\tau L-1}{\sqrt{3}}\right)
	+\frac{\pi}{6}\right]^{2} }.
\end{eqnarray}
At  the limit $\tau\rightarrow 0$, we have 
\begin{eqnarray}
\lim_{ \tau\rightarrow 0} E_n= \varepsilon
_n=\frac{\pi^2\hbar^2n^2}{2m_0L^2},
\end{eqnarray}
where $  \varepsilon_n$ is the spectrum of a free  particle in an infinite  square well potential of the basic quantum mechanics with the fundamental energy $  \varepsilon_1=\frac{\hbar^2\pi^2 }{2m_0 L^2}$. Thus, the energy
levels  can  be rewritten as
\begin{eqnarray} \label{enrg2}
	E_n=\frac{3}{4} \left[\frac{\tau L}{\arctan\left(\frac{2\tau L-1}{\sqrt{3}}\right)
		+\frac{\pi}{6} }\right]^2 \varepsilon_n<\varepsilon_n.
\end{eqnarray} 
Comparing this result (\ref{enrg2})  to various tandem of PDM systems \cite{25,27,30}, we realise that the parameters of deformations used in both approches have different  implications. Here the effects of the parameter $\tau$  induces  contractions of  quantum levels  which   consequently lead to a decrease in the amplitude of the  energy levels  while  the parameter of deformation in \cite{25,27,30} leads to an increase in the energy levels of the particle.

The generalized wave function and the probability density corresponding to  the energies (\ref{enrg1}) are given by
\begin{eqnarray}
\psi_n (x) &=&  A\sin\left(\frac{ n\pi }{\left[\arctan\left(\frac{2\tau L-1}{\sqrt{3}}\right)
	+\frac{\pi}{6}\right]}\left[\arctan\left(\frac{2\tau x-1}{\sqrt{3}}\right)
+\frac{\pi}{6}\right]\right),\label{wa}\\
\rho_n (x)& =&   A^2\sin^2\left(\frac{ n \pi }{\left[\arctan\left(\frac{2\tau L-1}{\sqrt{3}}\right)
	+\frac{\pi}{6}\right]}\left[\arctan\left(\frac{2\tau x-1}{\sqrt{3}}\right)
+\frac{\pi}{6}\right]\right).
\end{eqnarray}

At  the limit $\tau\rightarrow 0$, we have 
\begin{eqnarray}
\lim_{ \tau\rightarrow 0} \psi_n= \sqrt{\frac{2}{L}}\sin\left(\frac{n\pi}{L}x\right),\\
\lim_{ \tau\rightarrow 0} \rho_n= \frac{2}{L}\sin^2\left(\frac{n\pi}{L}x\right).
\end{eqnarray}
 \subsection{Numerical results and discussions}

 In this part, numerical results and discussions have been done to investigate the effect of the parameter $\tau$ on the amplitudes of  energy levels, on the wave functions and on the   probability densities.  For the calculations,  we set  $m=1$,\,\,$\hbar=1$
and $L=1$.

\begin{figure}[htbp]
	\resizebox{1\textwidth}{!}{
		\includegraphics{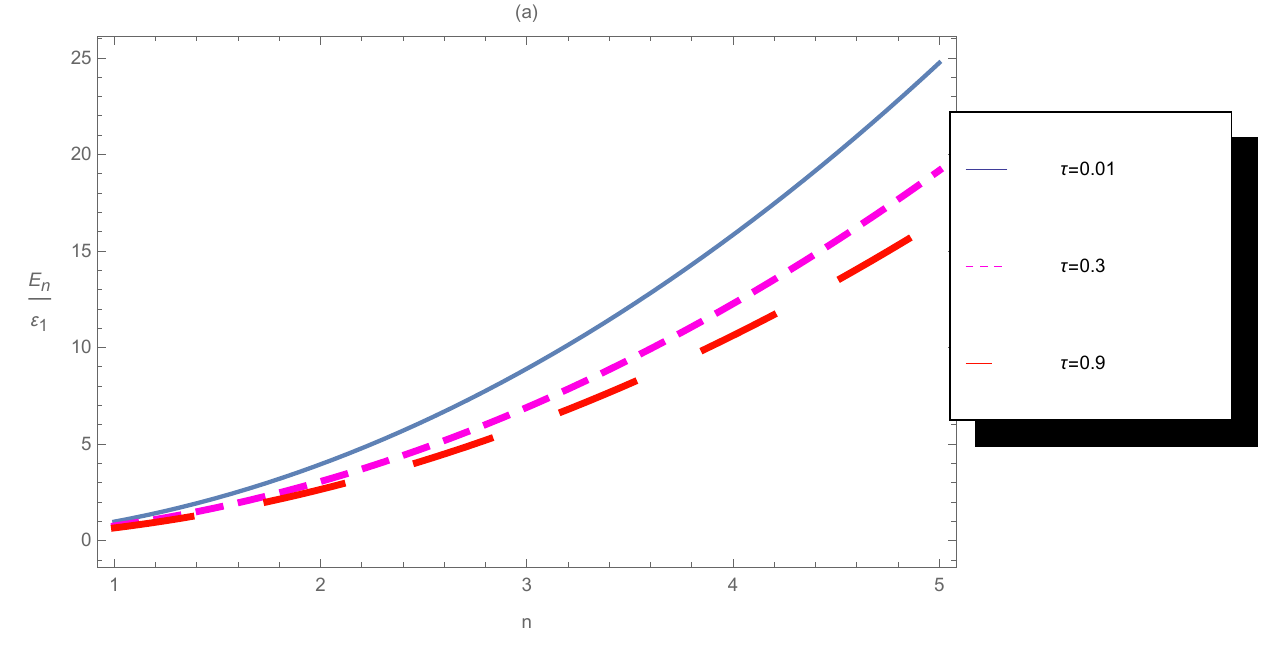}
	}
	\resizebox{1\textwidth}{!}{
		\includegraphics{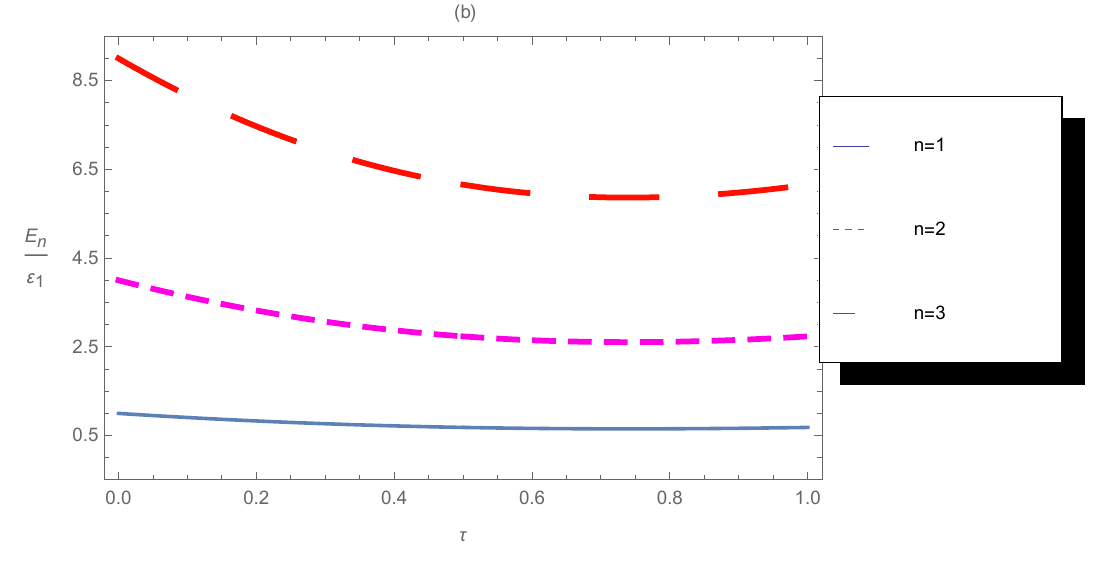}
	}
	\caption{\it \small The energy $E_n/\varepsilon_1$ of the particle in 1D box of length L = 1  with mass $m=1$ and $\hbar=1$. 
	}
	\label{figE}       
\end{figure}

Figure \ref{figE} illustrates the energy levels of the particle as  functions of   the quantum number $n$ and  the quantum gravity parameter $\tau$.   Figure \ref{figE}$a$ shows a plot of energy levels versus the quantum number $n$ for different values of $\tau$. Conversly to the graph obtained in  \cite{25,27,30}, one  observes that,  the amplitudes of   energy levels $E_n/\varepsilon_1$   decrease when $\tau$ increases. This is   confirmed by  the  analytical result obtained in  (\ref {enrg2}).  The same reasoning can be applied to  the  energy levels versus the parameter  $\tau$  for different values of $n$ (Figure \ref{figE}$b$). One   observes how the increase of   $\tau$ gradually deforms the energy levels as one increases the  quantum number $n$  from the  fundamental level. These  deformations curve the  quantum levels, allowing particle to jump from one state  to another with low energies. Moreover, the deformations observed at  these quantum levels are in perfect analogy with the theory of  GR  where the gravity  induces  deformation of the space.

\begin{figure}[htbp]
	\resizebox{0.52\textwidth}{!}{
		\includegraphics{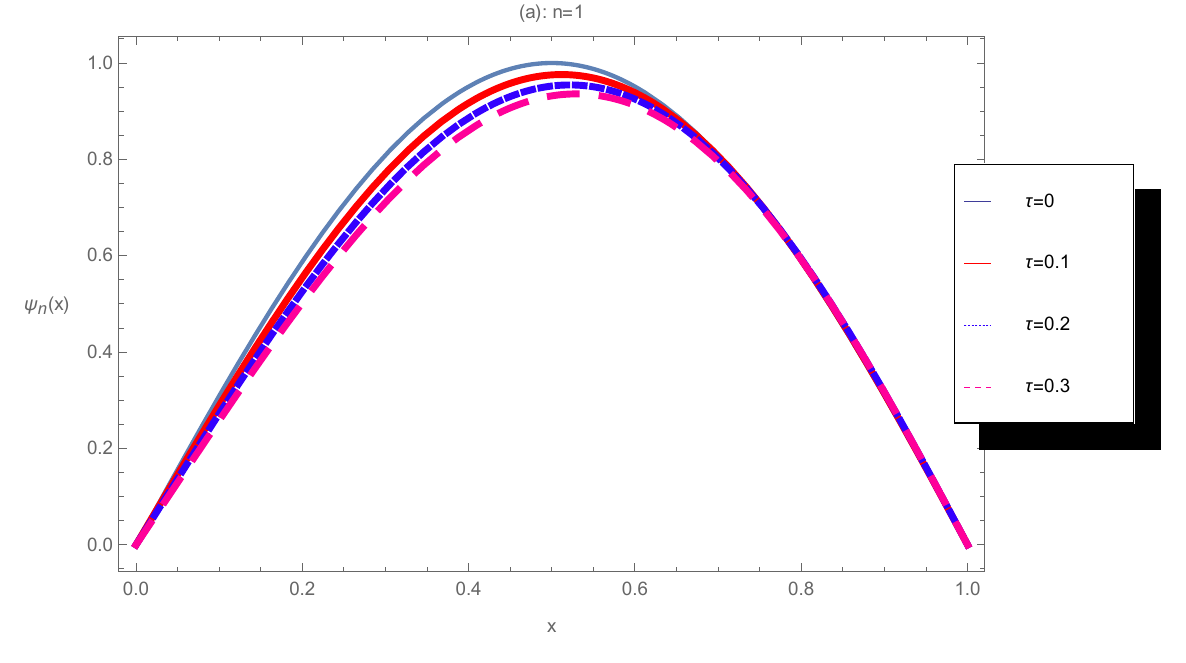}
	}
	\resizebox{0.52\textwidth}{!}{
		\includegraphics{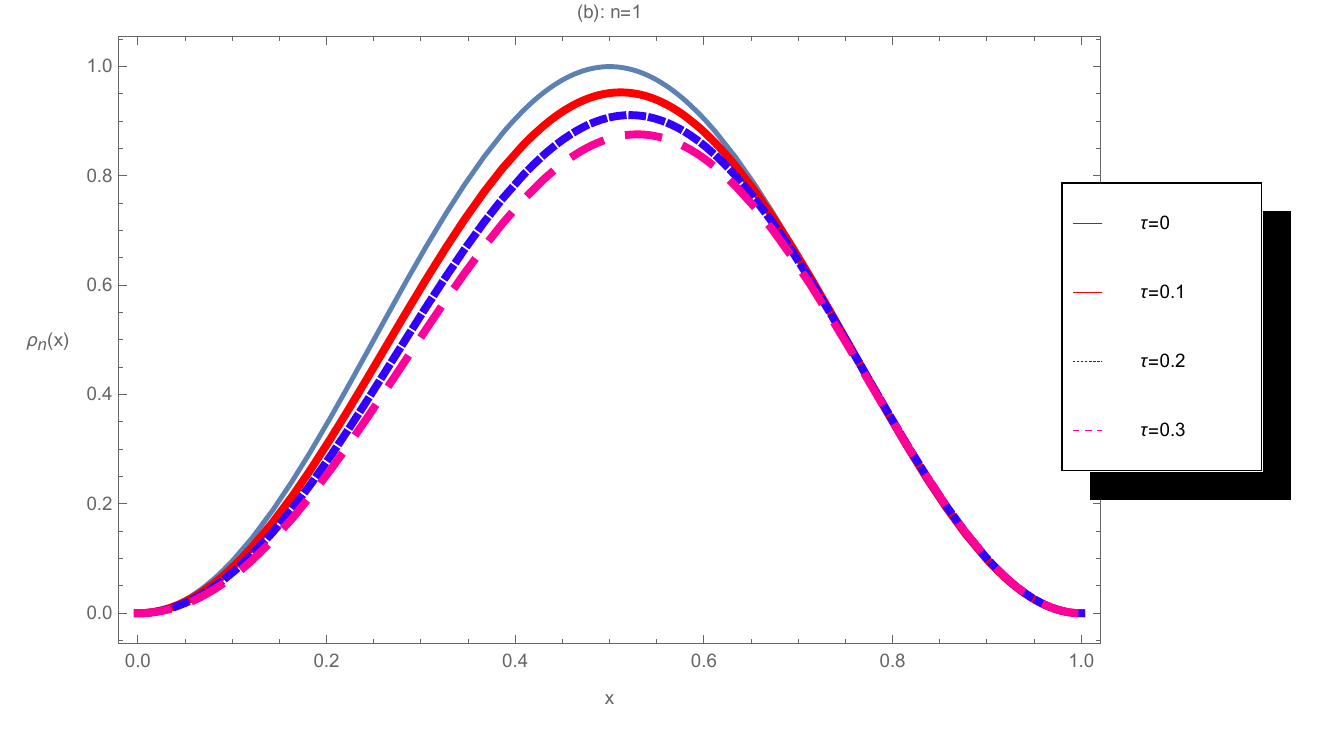}
	}
	\resizebox{0.52\textwidth}{!}{
		\includegraphics{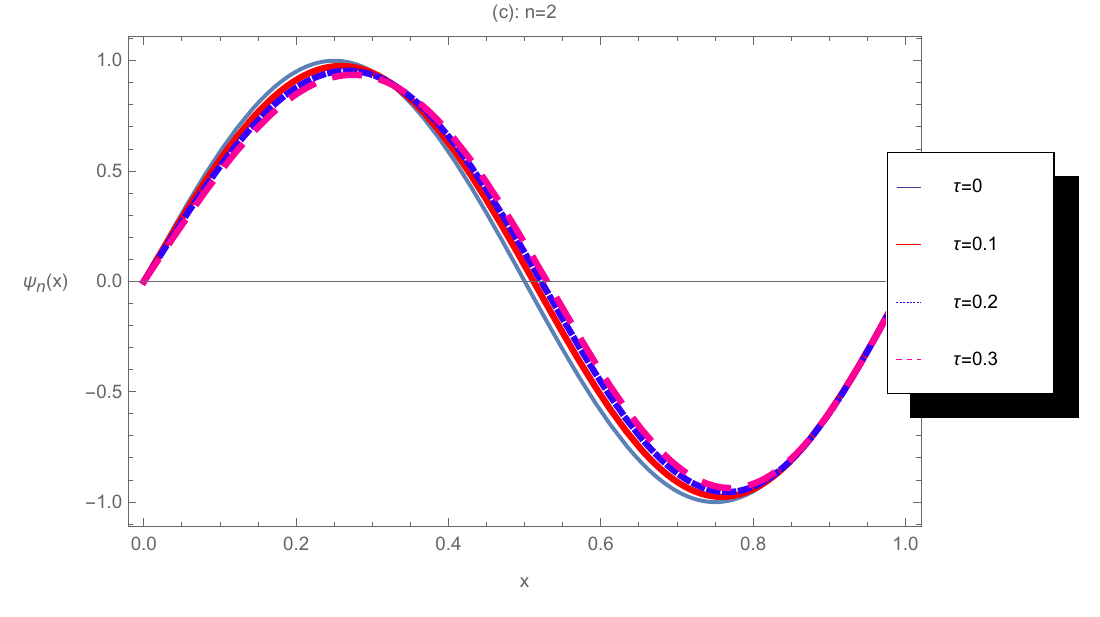}
	}
	\resizebox{0.52\textwidth}{!}{
		\includegraphics{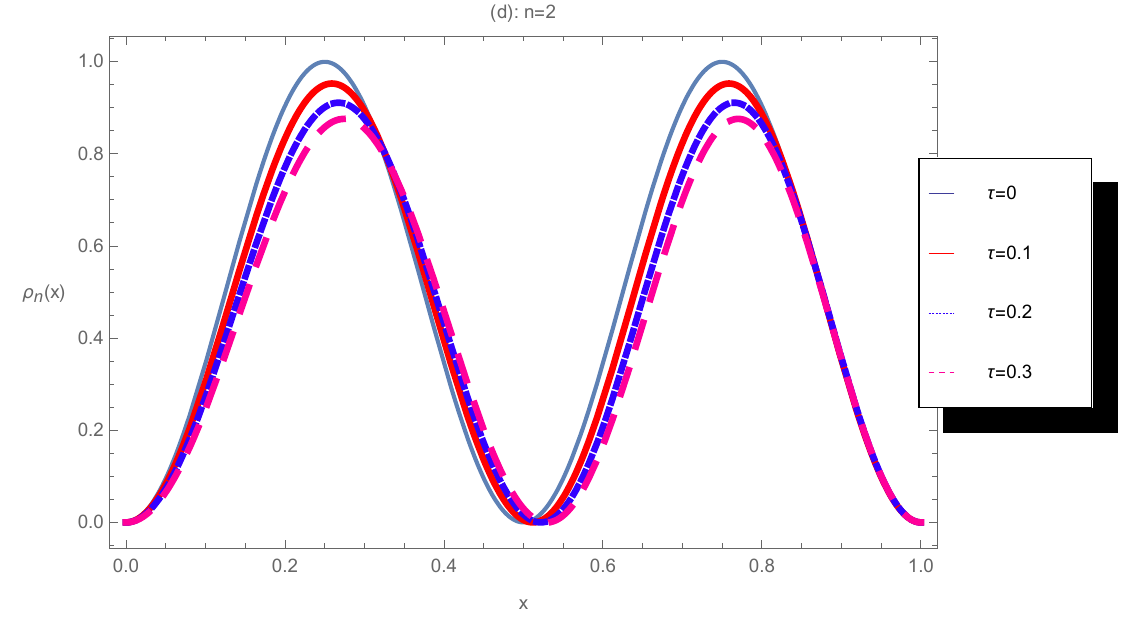}
	}
	\resizebox{0.52\textwidth}{!}{
		\includegraphics{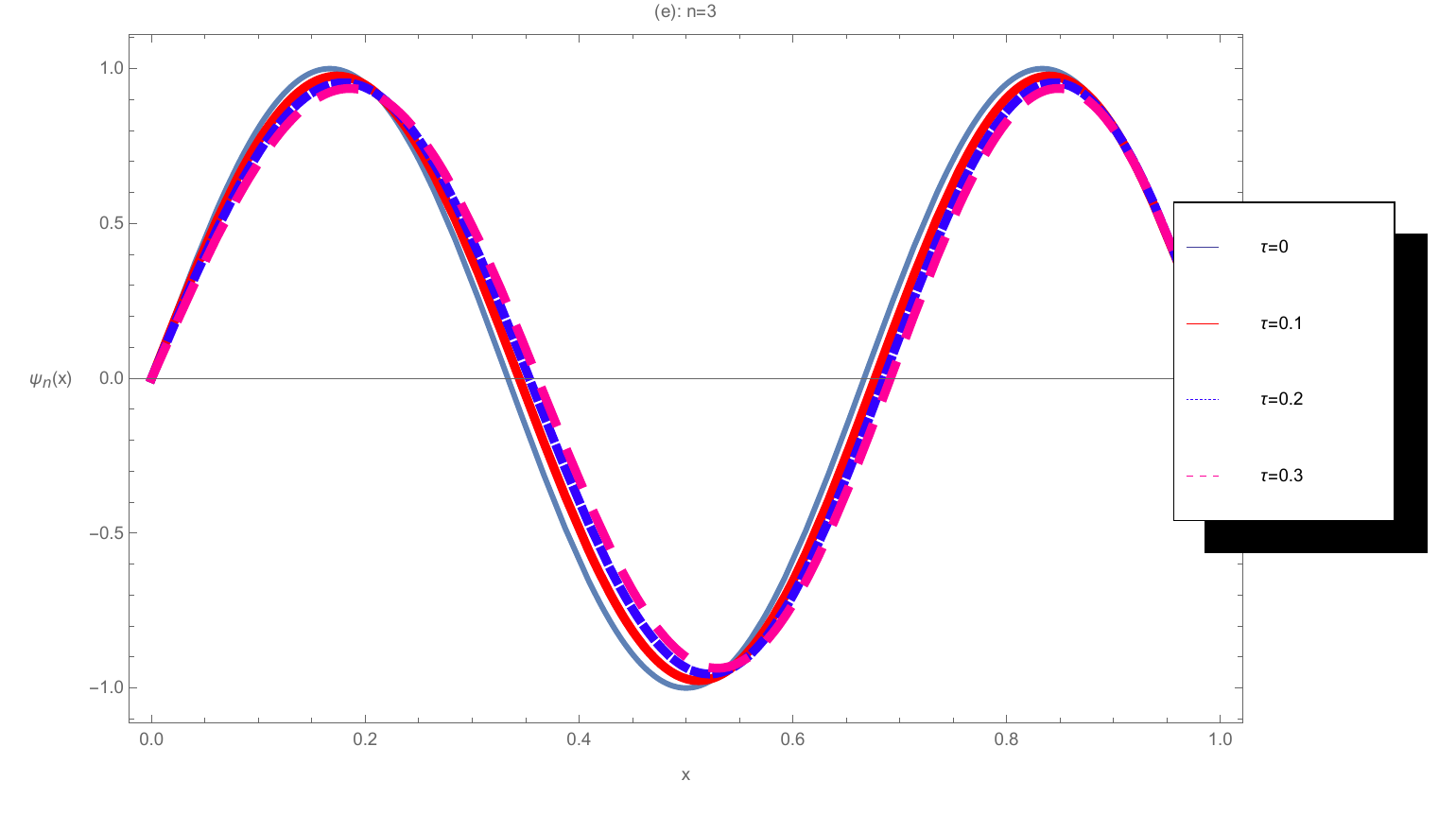}
	}
	\resizebox{0.52\textwidth}{!}{
		\includegraphics{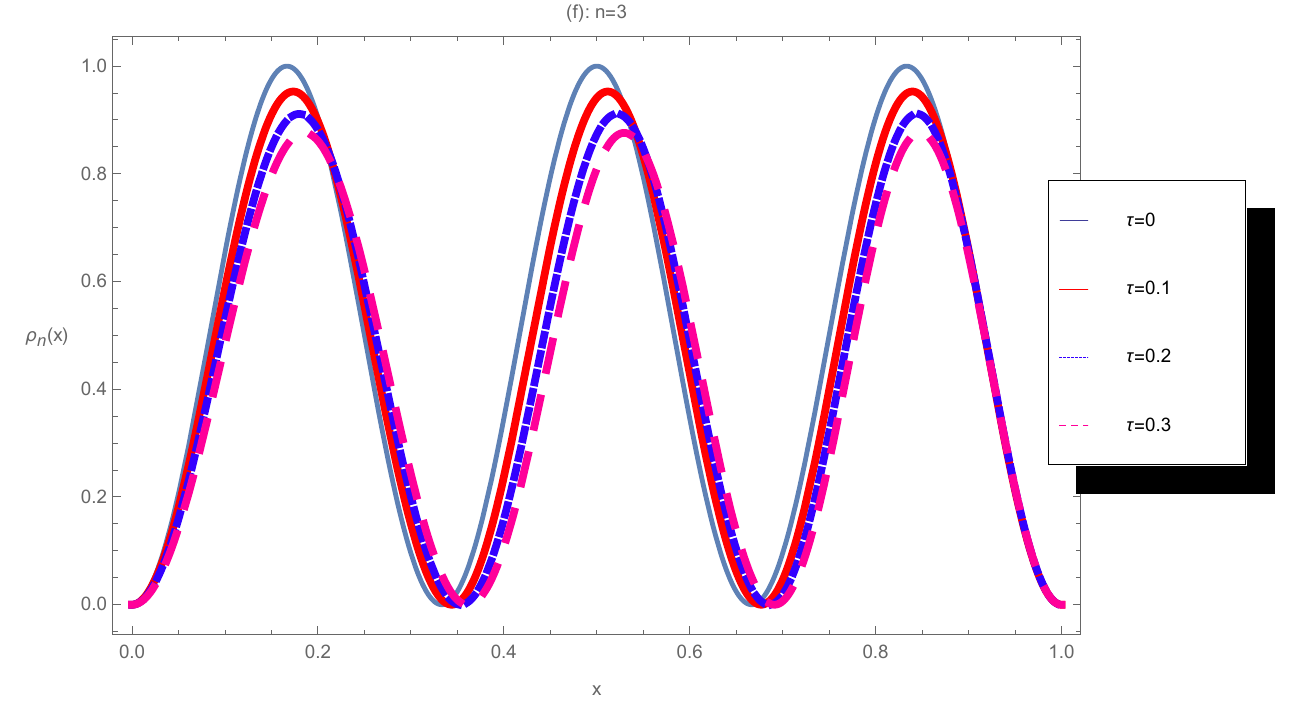}
	}
	\caption{\it \small Wave functions $\psi_n(x)$ (left column) and probability densities $\rho_n(x)$ (right column)  for a particle  confined in an infinite square well of length $L=1$ deformed by  the gravity parameter $\tau$. [(a) and (b)] $n = 1$ (ground state), [(c) and (d)] $n = 2$ (first excited state), [(e) and (f)] $n = 3$ (second excited state) 
	}
	\label{figwav}       
\end{figure}

Figure \ref{figwav} illustrates   the eigenfunctions $\psi_n(x)$  and   their respective probability densities $\rho_n(x)$ for the three  lowest states $n=1,n=2, n=3$  for some values of the  parameter $\tau \in\{0;0.1;0.2;0.3\}$. As expected, increase of  $n$  confines, shrinks and  contracts   $ \psi_n(x)$ and $\rho
_n(x)$. This fact comes to confirm the fundamental property of gravity which consists of length contraction. 

\begin{figure}[htbp]
	\resizebox{0.51\textwidth}{!}{
		\includegraphics{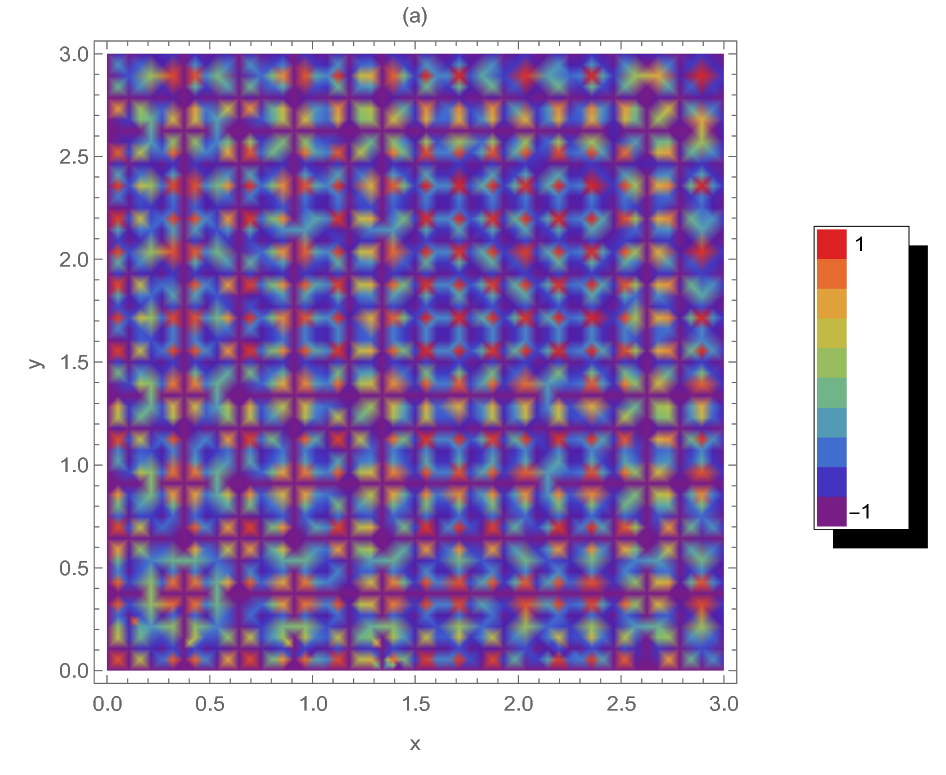}
	}
	\resizebox{0.51\textwidth}{!}{
		\includegraphics{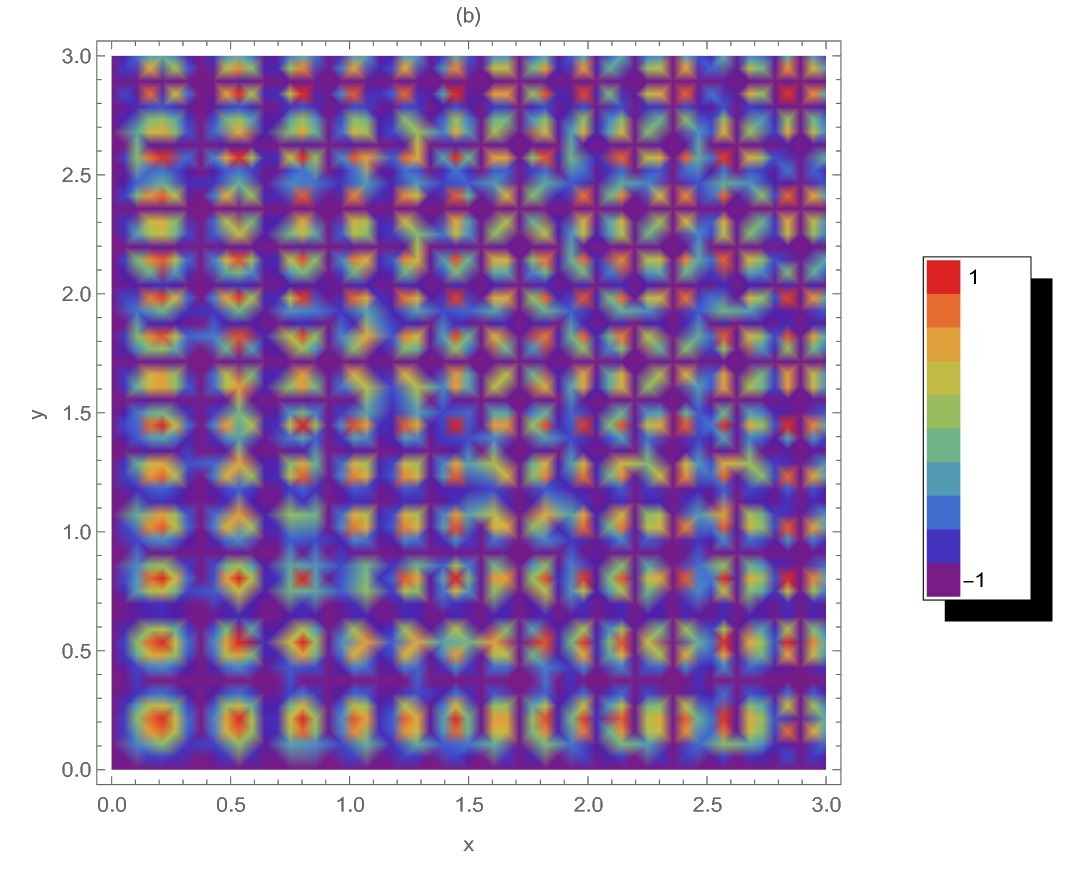}
	}
	\resizebox{0.51\textwidth}{!}{
		\includegraphics{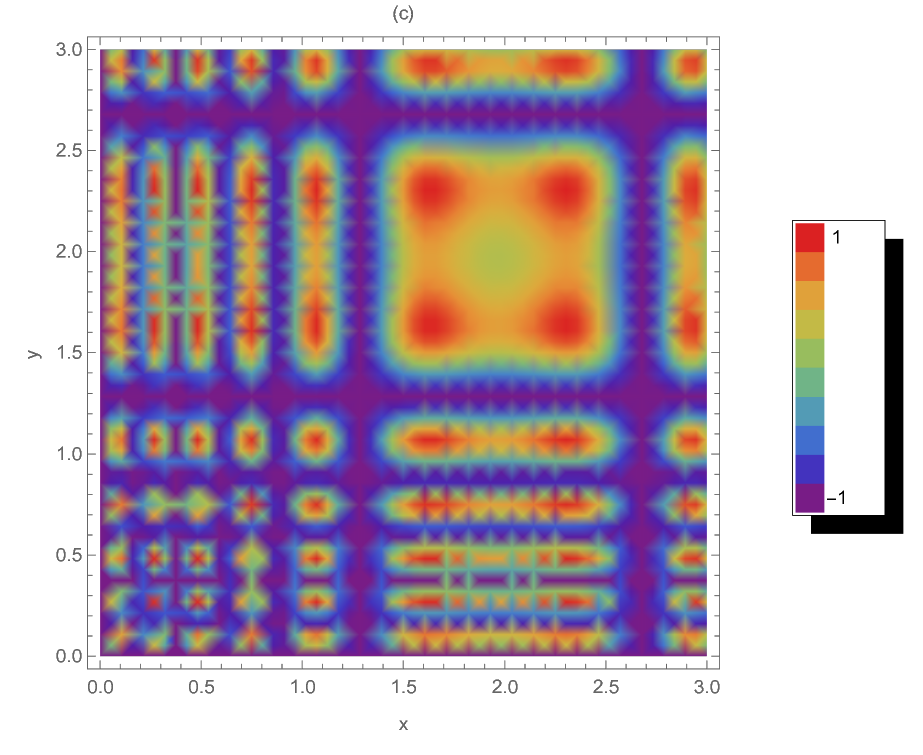}
	}
	\resizebox{0.51\textwidth}{!}{
		\includegraphics{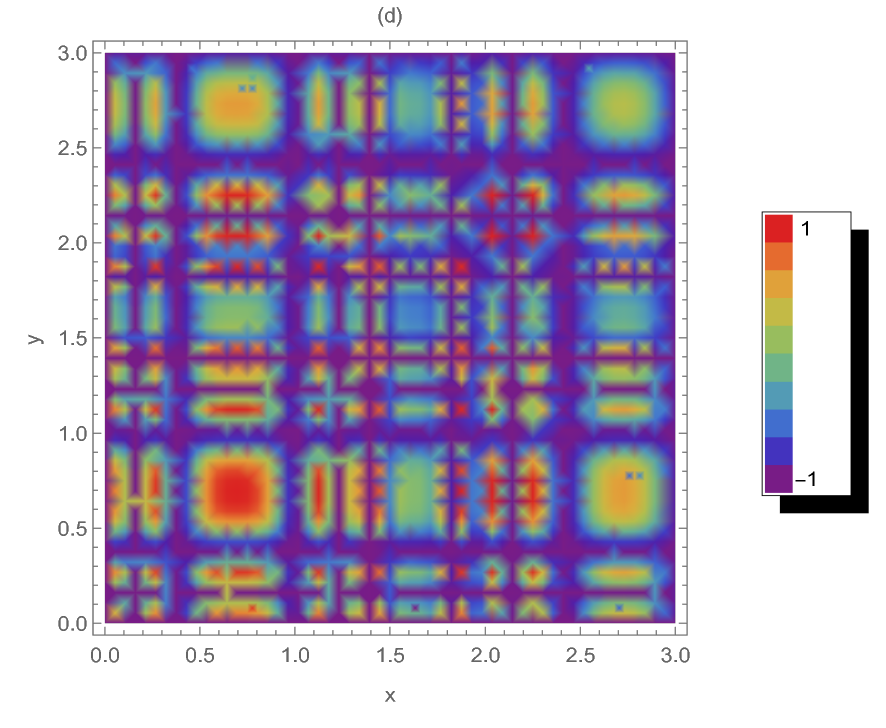}
	}
	\caption{\it \small The probability density of a two-dimensional infinite square well for (a) $n_x =n_y= 1$, (b) $n_x = n_y = 2$,
		(c) $n_x =n_y=3$, and (d) $n_x = n_y = 10$.
	}
	\label{figs}       
\end{figure}

The theory developed here can be easily
extended to two and three dimensions. For example, when
considering a square section of an infinite well, the corresponding wave function can be expressed as the product $\Psi_n(x,y)=\psi_n(x)\psi_n(y)$, where $ \psi_n(x)$ and $\psi_n(y)$ are the wave
functions in the $x$ and $y$ directions, respectively. 
The  probability density $\rho_n(x,y)$ of a particle moving in  a two-dimensional box is shown in Figure \ref{figs}  such  as  in  (a): $ n_x = n_y = 1$,
(b) $n_x = n_y = 2$,  (c) for $n_x = n_y = 3$ and in (d)  $n_x = n_y = 10$
where we have used $\tau = 0.1$  in all panels. Evidently, as pointed out through the    figure (\ref{figwav}),  it can be 
seen  that  for the first three lower  states the probability to
find a particle is practically the same everywhere in the square
well and this probability  strongly increases with the quantum numbers. But, 
unlike the figure reported in \cite{25,27}, here for large quantum numbers  $n_x = n_y = 10$,  one observes a decrease of probability densities.   

\section{Conclusion remarks} \label{sec5}
 In this paper, we study the dynamic of a position-dependent mass (PDM) trapped in an infinite  square well potential. The increase  of  quantum gravitational background fields of this well  has increased the PDM which  in returned has induced deformations of quantum energy levels. These deformations are more pronounced as one increases the quantum levels allowing the particle to jump from one state to another with low energies and with high probability densities. Furthermore, we have observed that the  states of this system  exhibit properties similar to Gaussian states of the  standard quantum mechanics which are  consequences of quantum gravitational fluctuations at this  scale.  As   expected, the increase of  quantum numbers $n$  has  confined, shrinked and contracted the  states of this system. This characterization  of quantum gravitational effects in this  well  is in perfect  analogy  with the one of general relativity (GR).

  Moreover,  it has been found that the mathematical foundations of the PDM systems are related to  some generalized algebraic structures   such as the $q$-deformed algebra \cite{25,47,48} and the   $\kappa$-deformed algebra \cite{49,50,51}. Thus, this version of a PDM system introduced in this paper could  give birth to a new algebraic structure namely the $\tau$-deformed algebra. This situation and the problem of  harmonic oscillator  could be  investigated in this  framework.

 \section*{Acknowledgments}
 The author acknowledges support from AIMS-Ghana
 under the Postdoctoral fellow/teaching assistance (Tutor) grant


\begin{thebibliography}{99}
	
\bibitem{1} L. Lawson, {\it Minimal and maximal lengths from 	position-dependent non-commutativity}, J. Phys. A: Math. Theor. {\bf 53},  115303 (2020) 


\bibitem{2} D. Amati, M. Ciafaloni and G. Veneziano, {\it Can Space-Time Be Probed Below the String Size?}, Phys.Lett. B {\bf 216}, 41-47 (1989)\\

\bibitem{3} C. Rovelli and L. Smolin, {\it  Discreteness of area and volume in quantum gravity}, Nucl. Phys. B. {\bf 442},  593–619  (1995). 


\bibitem{4} M. Maggiore, {\it A generalized uncertainty principle in quantum gravity}, Phys. Lett. B {\bf 304}, (1993) 65.

 \bibitem{5} M. Maggiore, {\it Quantum groups, gravity, and the generalized uncertainty principle}, Phys. Rev. D {\bf 49}, 5182 (1994) 

\bibitem{6} A. Kempf, G. Mangano and R. Mann, {\it Hilbert space representation of the minimal length uncertainty relation}, Phys. Rev. D {\bf 52}, 1108 (1995)

\bibitem{7} A. Kempf and  G. {\it Mangano, Minimal length uncertainty relation and ultraviolet regularization}, Phys. Rev. D. {\bf 55}, 7909-7920 (1997)

\bibitem{8} A. Kempf, {\it Uncertainty relation in quantum mechanics with quantum group symmetry}, J. Math. Phys. {\bf 35}, (1994) 4483-4496.

\bibitem{9} A. Fring, L. Gouba and F. Scholtz, {\it Strings from position-dependent noncommutativity}, J. Phys. A: Math. Theor. {\bf 43}, 345401 (2010)


\bibitem{10} L. Lawson,  L. Gouba  and G. Avossevou, {\it Two-dimensional noncommutative gravitational quantum well}, J. Phys. A: Math. Theor. {\bf 50}, 475202 (2017)

\bibitem{11} F. Scardiglia and R. Casadio, {\it Gravitational tests of the Generalized Uncertainty Principle},  Eur. Phys. J. C {\bf 75}, 425 (2015).

\bibitem{12} S. Das and E. Vagenas, {\it  Universality of Quantum Gravity Corrections}, Phys. Rev. Lett. {\bf 101}, 221301
 
\bibitem{13} G. Lambiase  and F. Scardigli, {\it Lorentz violation and generalized uncertainty principle}, Phys. Rev. D {\bf 97}, 075003 (2018)

\bibitem{14} F. Brau  and F. Buisseret,  {\it Minimal length uncertainty relation and gravitational quantum well}, Phys. Rev. D {\bf 74}, 0360002 (2006)

  \bibitem{15} G. Luciano, L. Petruzziello, {\it Generalized uncertainty principle and its implications on geometric phases in quantum mechanics}, Eur. Phys. J. Plus 136, 179 (2021)

\bibitem{16} G.  Luciano and L. Petruzziello, {\it GUP parameter from maximal acceleration}, Eur. Phys. J. C 79, 283 (2019)

\bibitem{17} L. Buoninfante, G. Lambiase, G.  Luciano and L.
Petruzziello, {\it Phenomenology of GUP stars}, Eur. Phys. J. C 80,
853 (2020)

\bibitem{18}  L. Buoninfante, G. Luciano and L. Petruzziello, {\it Generalized uncertainty principle and corpuscular gravity}, Eur. Phys. J. C 79, 663 (2019)


\bibitem{19} J. Schmöle, M. Dragosits, H. Hepach and
M. Aspelmeyer, {\it A micromechanical proof of principle
experiment for measuring the gravitational 	force of milligram masses},
Class. Quantum Grav. {\bf 33}, 125031  (2016) 

\bibitem{20} I. Pikovski, M. R. Vanner, M. Aspelmeyer, M. Kim, and C.
Brukner {\it Probing Planck-scale physics with quantum optics},  Nat. Phys. {\bf8}, 393 (2012)

\bibitem{21} M. Bawaj, C. Biancofiore et al.{Probing deformed commutators with macroscopic harmonic oscillators}, Nat Commun {\bf 6}, 7503 (2015)

\bibitem{22} D. Gao and M. Zhan, {\it Constraining the generalized uncertainty principle with cold atoms},  Phys. Rev. A 94, 013607 (2016).

\bibitem{23}   L. Perivolaropoulos, {\it Cosmological horizons, uncertainty principle, 	and maximum length quantum mechanics}, Phys. Rev. D {\bf 95}, 103523 (2017)

 \bibitem{24} Oldwig von Roos, {\it Position-dependent effective masses in semiconductor theory}, Phys Rev B {\bf 27}, 12 (1983)
 
\bibitem{25} {\color{red}R. N. Costa Filho}, M. Almeida, G. Farias, and J. Andrade Jr.,     {\it Displacement operator for quantum systems with position-dependent mass},
Phys Rev A, {\bf 84}, 050102 (2011)

\bibitem{26} B. da Costa and  E. Borges, {\it Generalized space and linear momentum operators  	in quantum mechanics}, J. Math Phys  {\bf 55}, 062105 (2014)

\bibitem{27} S. Habib Mazharimousavi, {\it Revisiting the displacement operator for quantum systems with position-dependent mass},  Phys Rev A,  {\bf  85}, 034102 (2012)

\bibitem{28} G. Bruno  da Costa  and Ernesto P. Borges,  {\it A position-dependent mass harmonic oscillator   	and deformed space}, J. Math. Phys. {\bf 59}, 042101 (2018)

\bibitem{29}  M. Rego-Monteiro and F. Nobre, {\it Classical field theory for a non-Hermitian Schrodinger equation with position-dependent masses},  Phys Rev A {\bf 88}, 032105 (2013)


\bibitem{30} B. da Costa, I. Gomez, and M. Portesi, {\it $\kappa$-Deformed quantum and classical mechanics  	for a system with position-dependent
	effective mass}, J. Math. Phys. {\bf 61}, 082105 (2020)

\bibitem{31} G. Bastard, {\it Wave mechanics applied  to semiconductor heterostructures}, Editions de Physique, Les Ulis, France, 1988)

\bibitem{32} G.  Barbagiovannia, D. Lockwoodb, R. N. Costa Filho, L.
Goncharovad and P. Simpsond, {\it Quantum confinement in Si and Ge nanostructures: Effect of crystallinity}, Proc. SPIE 8915, Photonics North (2013), 891515 DOI: 10.1117/12.2036323

 \bibitem {33}  E.  Barbagiovanni, S. Cosentino, D.  Lockwood, R. N. Costa Filho,  A. Terrasi, and S. Mirabella, {\it Influence of interface potential on the effective mass in Ge nanostructures},  Journal of Applied Physics, {\bf 117}, 154304 (2015)
 
 \bibitem{34} R. Szabo, {\it Quantum field theory on noncommutative spaces}, Phys.Rept. {\bf 378},  207 (2003)
  
 \bibitem{35} F. Delduc, Q. Duret, F. Gieres, M. Lefrancois, {\it Magnetic fields in 	noncommutative quantum mechanics},  J. Phys: Conf. Ser. {\bf 103}, 012020 (2008)
  
 \bibitem{36}  Richard J. Szabo, {\it Quantum field theory on noncommutative spaces} Physics Reports Volume 378, Issue 4, May 2003, Pages 207-299
 
 \bibitem{37} D.  Bigatti  and S. Susskind, {\it Magnetic fields, branes and noncommutative geometry} Phys. Rev. D {\bf  62}, 066004  (2000)
 
 \bibitem{38} C. Bender,{\it  Making sense of non-Hermitian Hamiltonians}, Rept. Prog. Phys. {\bf 70}, 947-1018 (2007)
 
 \bibitem{39} A. Mostafazadeh, {\it Pseudo-Hermiticity versus PT symmetry. The necessary condition for the reality of the spectrum}, J. Math. Phys. {\bf 43},  205-214 (2002)
 
 \bibitem{40} F. Scholz, H. Geyer, and F. Hahne, {\it Quasi-Hermitian operators in quantum mechanics and variational principle}, Ann. Phys. {\bf 213}, 74 (1992)
 
 \bibitem{41} J. Dieudonné, {\it Quasi-Hermitian operators, Proceedings of the International Symposium on Linear Spaces}, Jerusalem 1960, Pergamon, Oxford,  115-122 (1961)
 
 \bibitem{42} M. Froissart, {\it  Covariant formalism of a field with indefinite metric}, II Nuovo Cimento {\bf 14}, 197-204 (1959)
 
 \bibitem{43} E. Sudarshan, {\it Quantum Mechanical Systems with Indefinite Metric. I}, Phys. Rev. {\bf 123},  2183-2193 (1961)
 
 \bibitem{44} A. Mostafazadeh, {\it Pseudo-Hermiticity versus PT symmetry. The necessary condition for the reality of the spectrum}, J. Math. Phys. {\bf 43},  205-214 (2002).
 
 \bibitem{45} D. Gao and M. Zhan, {\it Constraining the generalized uncertainty principle  with cold atoms}, Phys. Rev. A 94, 013607 (2016).
 
 \bibitem{46} Jonas Schmöle, Mathias Dragosits, Hans Hepach and
 Markus Aspelmeyer, {\it A micromechanical proof-of-principle
 	experiment for measuring the gravitational 	force of milligram masses},
 Class. Quantum Grav. {\bf 33}, 125031  (2016)
 
 \bibitem{46'} K. Nozari  and A. Etemadi,  {\it  Minimal length, maximal momentum and Hilbert space representation
 of quantum mechanics}, Phys. Rev. D  {\bf 85}, 104029  (2012)
 
\bibitem{47} E. Borges, {\it A possible deformed algebra and calculus
	inspired in nonextensive thermostatistics}, Physica A {\bf 340}, 95 (2004).

\bibitem{48} C. Tsallis, {\it Introduction to Nonextensive Statistical Mechanics: Approaching a Complex World}, (Springer, New York, 2009)

\bibitem{49} G. Kaniadakis and A. Scarfone, {\it A new one-parameter deformation of the exponential function}, Physica A {\bf 296}, 405 (2001).
\bibitem{50} G. Kaniadakis, {\it Statistical mechanics in the context of special relativity}, Phys. Rev. E {\bf 66}, 056125 (2002).
\bibitem{51} G. Kaniadakis, {\it Statistical mechanics in the context of special relativity. II} Phys. Rev. E {\bf 72}, 036108 (2005)


\end{thebibliography}
\end{document}